\documentclass[prl,10pt,superscriptaddress,twocolumn]{revtex4}

\usepackage[latin1]{inputenc}
\usepackage[english]{babel}
\usepackage[draft]{hyperref}
\usepackage{graphicx}
\usepackage{amsmath}
\usepackage{color}
\usepackage{float}

\begin{document}

\title{Quantum Enhancement of the Zero-Area Sagnac Interferometer Topology for Gravitational Wave Detection}

\author{Tobias Eberle}
\affiliation{Max-Planck-Institut für Gravitationsphysik
(Albert-Einstein-Institut) and\\ Institut für Gravitationsphysik
der Leibniz Universität Hannover, Callinstraße 38, 30167 Hannover,
Germany} \affiliation{Centre for Quantum Engineering and
Space-Time Research - QUEST, Leibniz Universität Hannover,
Welfengarten 1, 30167 Hannover, Germany}
\author{Sebastian Steinlechner}
\author{Jöran Bauchrowitz}
\author{Vitus Händchen}
\author{Henning Vahlbruch}
\affiliation{Max-Planck-Institut für Gravitationsphysik
(Albert-Einstein-Institut) and\\ Institut für Gravitationsphysik
der Leibniz Universität Hannover, Callinstraße 38, 30167 Hannover,
Germany}
\author{Moritz Mehmet}
\affiliation{Max-Planck-Institut für Gravitationsphysik
(Albert-Einstein-Institut) and\\ Institut für Gravitationsphysik
der Leibniz Universität Hannover, Callinstraße 38, 30167 Hannover,
Germany} \affiliation{Centre for Quantum Engineering and
Space-Time Research - QUEST, Leibniz Universität Hannover,
Welfengarten 1, 30167 Hannover, Germany}
\author{Helge Müller-Ebhardt}
\author{Roman Schnabel}
\email[corresponding author: ]{roman.schnabel@aei.mpg.de}
\affiliation{Max-Planck-Institut für Gravitationsphysik
(Albert-Einstein-Institut) and\\ Institut für Gravitationsphysik
der Leibniz Universität Hannover, Callinstraße 38, 30167 Hannover,
Germany}

\begin{abstract}
Only a few years ago, it was realized that the zero-area Sagnac interferometer topology is able to perform quantum nondemolition measurements of position changes of a mechanical oscillator. Here, we experimentally show that such an interferometer can also be efficiently enhanced by squeezed light. We achieved a nonclassical sensitivity improvement of up to $8.2$~dB, limited by optical loss inside our interferometer. Measurements performed directly on our squeezed-light laser output revealed squeezing of $12.7$~dB. We show that the sensitivity of a squeezed-light enhanced Sagnac interferometer can surpass the standard quantum limit for a broad spectrum of signal frequencies without the need for filter cavities as required for Michelson interferometers. The Sagnac topology is therefore a powerful option for future gravitational-wave detectors, such as the Einstein Telescope, whose design is currently being studied.
\end{abstract}

\maketitle

%\section{Introduction}
All currently operating interferometric gravitational-wave (GW)
detectors (LIGO~\cite{SHOEMAKER2004}, VIRGO~\cite{ACERNES2002},
GEO~\cite{WIL2002} and TAMA~\cite{ANDO2001}) are Michelson
interferometers. Their purpose is to measure the position changes (displacements) of
quasi free-falling mirrors thereby revealing changes of space-time
curvature, i.e.\,gravitational waves. The current
detectors are aiming for the first direct observation of
gravitational waves. Future detectors will aim for establishing
gravitational wave astronomy which will require a considerable
increase of the detectors' displacement sensitivity. The Einstein
Telescope (ET)~\cite{ET2009} is an on-going European design study
project for such a gravitational wave detector. An important issue
for future detectors is a reduction of the quantum measurement
noise (photon shot-noise) and quantum back-action noise (quantum fluctuations in the radiation pressure acting on the mirrors) for a given laser power. This way, the standard quantum limit (SQL) can eventually be surpassed allowing for
quantum-non-demolition (QND) displacement measurements. Future detectors will most likely be operated at cryogenic
temperatures in order to reduce thermally excited motions of the
mirror surfaces and optical absorption will set an upper limit to
the laser power inside the interferometer. For Michelson
interferometers a nonclassical reduction (squeezing) of the
quantum measurement noise can be
achieved by injecting squeezed light \cite{Caves81,McKenzie2002,Vahlbruch2005,Goda2008}. 
Squeezing of
back-action noise is also possible but turned out to be
experimentally challenging because in Michelson interferometers
the back-action noise depends
on Fourier frequency.  Long-baseline narrow-band filter cavities are
required to compensate for the frequency dependence
\cite{KLMTV01} which complicates the topology and introduces additional optical loss. 

The Sagnac interferometer was originally invented by G.\,Sagnac in
1913 \cite{Sagnac1913} and can be used as a rotation sensor. However, the Sagnac
interferometer is also sensitive to displacements of its mirrors if
the latter are not located at half the round trip length. By
setting the area that is enclosed by the two counter-propagating
beams to zero the interferometer can be made insensitive to
rotations though keeping it sensitive for displacements.
First experimental tests of zero-area Sagnac interferometers for
gravitational wave detection were performed in the late
1990s~\cite{Sun1996,Shaddock1998}. No advantage in comparison with the Michelson interferometer
could be found. But in 2003,
Chen~\cite{Chen03} realized that a zero-area Sagnac interferometer, 
by its very nature, can suppress the quantum back-action noise for displacement
measurements over a broad frequency band. 
Since then the
Sagnac topology for gravitational wave detection has not been
further experimentally investigated. It has also not been
theoretically investigated if the Sagnac's QND property is
compatible with squeezing of its quantum measurement noise.

In this Letter, we experimentally demonstrate that a Sagnac
interferometer can be efficiently enhanced by the injection of
squeezed light. We achieved, to the best of our knowledge, the
strongest squeezing of quantum measurement noise in an
interferometric device ever observed. We also present a
theoretical analysis by how much the sensitivity of a possible
design of the Einstein Telescope can surpass the standard quantum
limit when realistic values for squeezing factor and detection loss
apply.

%\section{Experiment}

\begin{figure}[ht]
  \includegraphics[width=8.6cm]{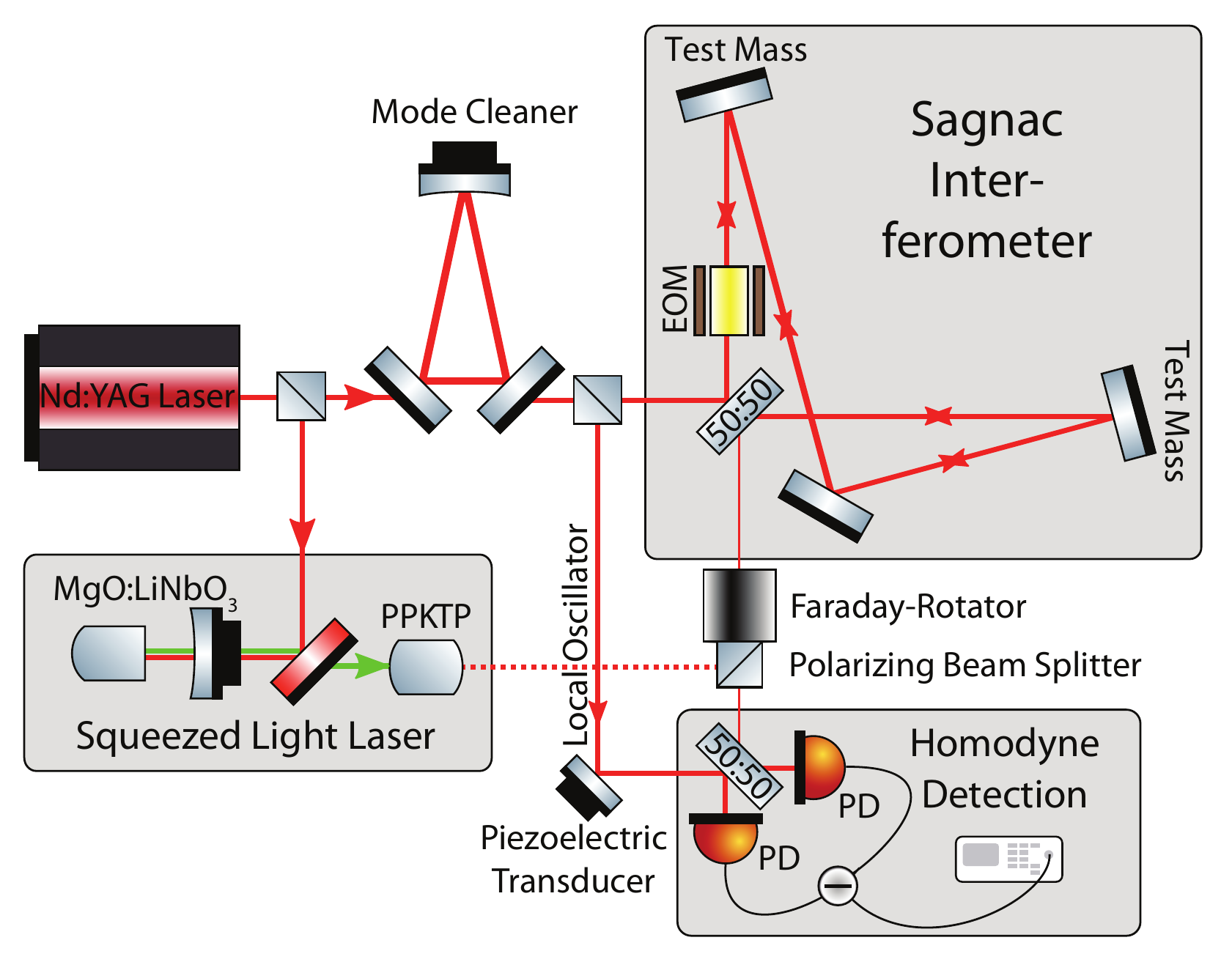}
  \caption{Schematic of the experiment. A continuous-wave 
  laser beam at 1064\,nm was used to produce squeezed light, 
  operate a zero-area Sagnac interferometer and perform the 
  interferometer readout with balanced homodyne detection. 
  The squeezed field was injected into the interferometer 
  through its dark port by means of a polarizing beam splitter and a Faraday rotator. %EOM: electro-optical modulator, 
  %MC: spatial mode cleaner cavity
  }
  \label{fig:experiment}
\end{figure}

Figure~\ref{fig:experiment} shows the schematic of our experiment.
The Sagnac interferometer had a round trip length of $50$~cm and
was set up such that it was insensitive to rotations, i.e.\,with an
effective zero area. A continuous-wave laser beam at a wavelength of $1064$~nm was used to generate squeezed light, and also provided the carrier 
field of the Sagnac interferometer and the local oscillator of the 
balanced homodyne detector. An electro-optical phase modulator (EOM) was placed in one interferometer arm
right after the beam splitter in order to generate an optical
displacement signal similar to that of a gravitational wave.

Into the second port of the Sagnac interferometer we injected a
squeezed mode of light, precisely mode-matched to the
interferometer mode. The
interferometer signal and the squeezed quantum noise were
detected by balanced homodyne detection by means of a pair of high
quantum efficiency InGaAs photo diodes. The squeezed light was
produced by parametric down-conversion in periodically poled
potassium titanyl phosphate (PPKTP). The end faces of the $8.9$~mm
long bi-convex crystal were dielectrically coated in order to form
a monolithic cavity. One surface was highly reflecting for both
laser wavelengths involved, the fundamental at $1064$~nm and the
pump at $532$~nm. The second surface had a reflectivity of
$R=90$~$\%$ for the fundamental and $R=20$~$\%$ for the pump
field. To achieve phase matching the crystal was temperature
controlled to $37.8$~$^\circ$C. 
% The laser's frequency was tuned to meet a resonance of the monolithic cavity. 
The required
pump beam for the parametric process was generated by
second-harmonic generation in $7$~$\%$ magnesium-doped lithium
niobate.

%\section{Results}

\begin{figure}[ht]
  \includegraphics[width=8.2cm]{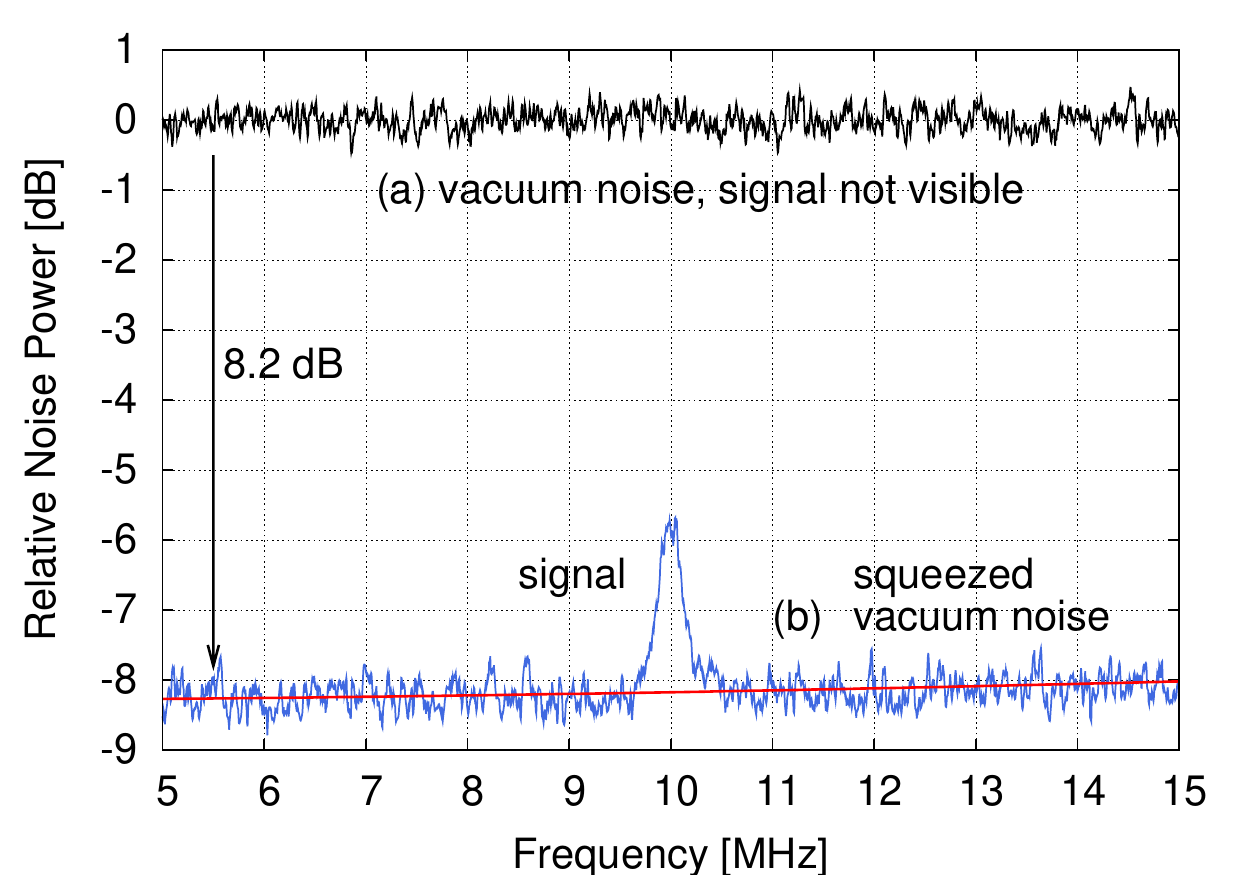} 
  \caption{$10$~MHz phase modulation measured with the Sagnac interferometer without (trace (a)) and with (trace (b)) squeezed light input. Trace (a) corresponds to the vacuum-noise, is normalized to unity, and completely buries the signal. Both traces are averaged twice, each measured with a resolution bandwidth of $300$~kHz and a video bandwidth of $300$~Hz. The homodyne detector dark noise was $22$ to $25$\,dB below the vacum noise and was subtracted from the data. The solid line shows a model for the squeezing strength fitted to the data.}
  \label{fig:resultssagnac}
\end{figure}

Figure~\ref{fig:resultssagnac} shows the squeezing of quantum
measurement noise in our zero-area Sagnac interferometer. The
upper trace (a) corresponds to the vacuum fluctuations entering 
the interferometer through its signal port when the squeezed light was blocked,
and was normalized to unity. The lower trace (b) shows the noise
floor when squeezed light was injected into the interferometer's
signal port. The nonclassical noise reduction was $8.2$~dB. For \textit{both} traces a $10$~MHz signal was
applied using the electro-optical modulator inside the
interferometer. In the case of trace (a) this signal is not
visible, however clearly detected in trace (b). The light power circulating in the
Sagnac interferometer was $570$~$\mu$W. The contrast of the
interferometer was measured to $\mathcal{C} = 99.7$~$\%$. The
visibility at the Sagnac's beam splitter for the injection of
squeezed light was $99.8$~$\%$ and the visibility at the homodyne
detectors' beam splitter was $99.7$~$\%$. The local oscillator
power was $20$~mW and the green pump power was $80$~mW.

\begin{figure}[ht]
  \includegraphics[width=8.2cm]{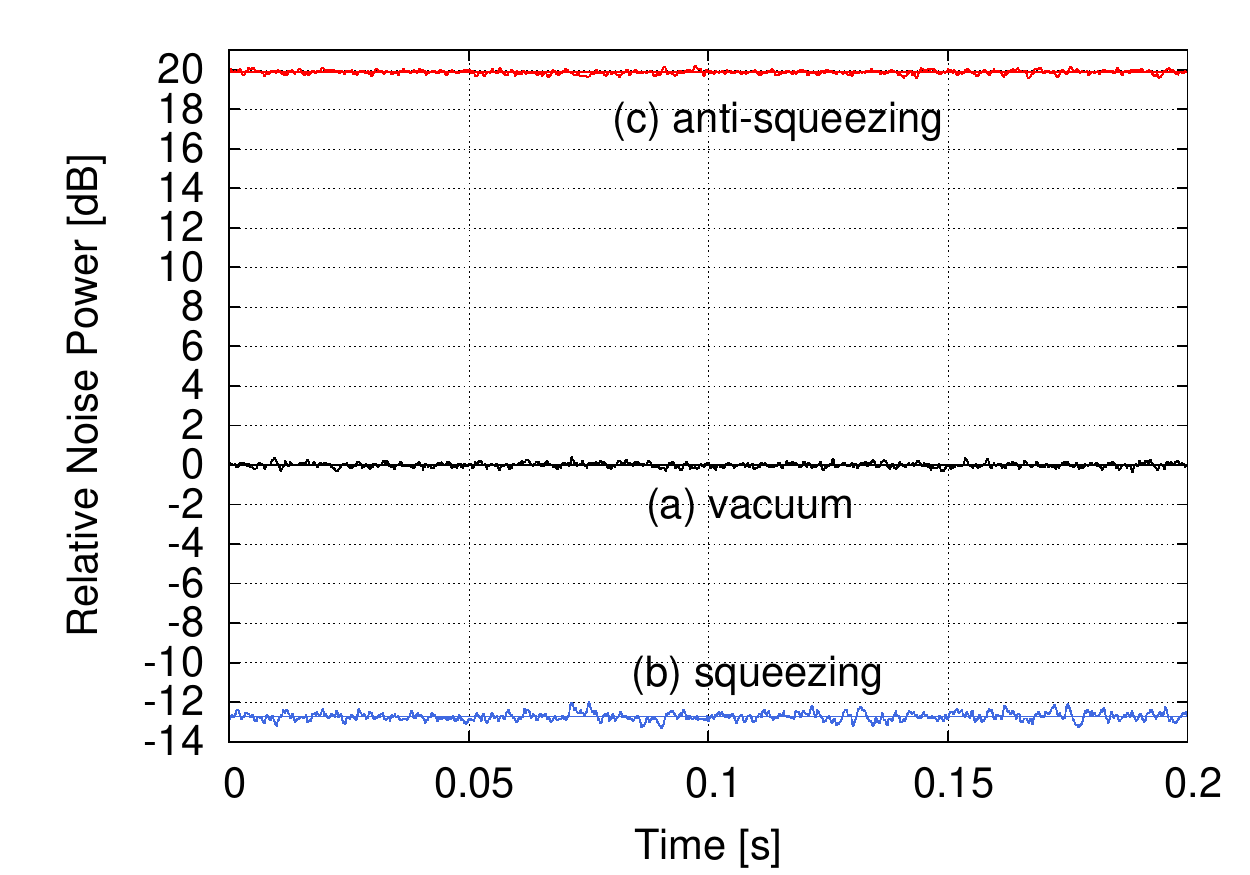}
  \caption{Characterization of our squeezed light laser at a sideband frequency of $5$~MHz. All traces are normalized to the vacuum noise reference of the homodyne detector's local oscillator beam (a). Trace (b) shows the noise in the squeezed field quadrature; trace (c) shows the antisqueezing in the orthogonal quadrature. %The curves are averaged twice and recorded with a resolution bandwidth of $300$~kHz and a video bandwidth of $300$~Hz. 
  Detector dark noise is at -25\,dB and, here, \textit{not} subtracted from the data.}
  \label{fig:sqz}
\end{figure}

Figure~\ref{fig:sqz} characterizes of the squeezed
light laser output without the optical loss introduced by the
Sagnac interferometer. %The second harmonic pump power was again $80$~mW. 
Trace (a) shows the vacuum noise measured with the squeezed
light blocked. Trace (b) and (c) represent the noise in the squeezed
 and antisqueezed
quadrature, respectively. 
%The detector dark noise was $25$~dB to the vacuum noise and not subtracted from the data. 
We observed $12.7$~dB of squeezing and $19.9$~dB
of antisqueezing. To the best of our knowledge the squeezing factor of $12.7$~dB represents the
highest value ever observed. Previously, squeezing factors above 10 were observed in Refs.~\cite{Vahlbruch08,Moritz11dB}.
The squeezed ($s$) and antisqueezed ($as$) quadrature variances observed in
Figs.~\ref{fig:resultssagnac} and \ref{fig:sqz} can be described
by the following equation~\cite{Takeno2007}
\begin{equation}
\label{eq:sqz_var}
  V_{s,as} = 1\pm \eta\gamma\frac{4\sqrt{P_{532}/P_\text{th}}}{(1\mp\sqrt{P_{532}/P_\text{th}})^2 + 4 K(f)^2}\ ,
\end{equation}
where the total optical loss is described by the detection
efficiency $\eta$ and the nonlinear cavity escape efficiency
$\gamma$. $P_{532}$ is the parametric pump power, $P_\text{th}$ the
threshold power and $K(f)=2\pi f / \kappa$, the ratio between the
Fourier frequency $f$ and the cavity decay rate $\kappa$. The
decay rate is defined by $\kappa = (T+L)c/(n\cdot l)$ with output
coupler transmission $T=0.1$, the intra-cavity loss $L$, the speed
of light in vacuum $c$, the refractive index of PPKTP $n=1.83$ and
the round trip length $l=2\cdot 8.9$~mm. All the variances
observed can consistently be described by setting
$P_{532}/P_{th}=2/3$ and $L=3.56\cdot 10^{-3}$ in Eq.~(\ref{eq:sqz_var}). For the
measurements in Fig.~\ref{fig:resultssagnac} we deduce a total
optical loss of $14$~$\%$ ($\eta\gamma = 0.86$). For the squeezed light
characterization in Fig.~\ref{fig:sqz} we deduce a value of $4.5$~$\%$
($\eta\gamma = 0.955$) which also accounts for about $1$~$\%$ loss of our nonperfect photo-diodes. 
%Setting the total loss to zero, Eq.~\ref{eq:sqz_var} infers squeezing and anti-squeezing factors of $\pm 20.1$~dB. 
%
The loss of about $10$~$\%$ introduced by the
Sagnac interferometer, was in good agreement with independent loss
measurements. First of all, the squeezed light was passed twice
through a Faraday rotator (Fig.~\ref{fig:experiment}) which
introduced optical loss of about $4$~$\%$. About $1$~$\%$ of the light was
transmitted through the Sagnac interferometer because its central
beam splitter deviated from its optimum $50:50$ splitting ratio.
About $1.5$~$\%$ loss was due to the transmission through the EOM crystal and the imperfect antireflection coating of the second beam splitter
surface. And finally each of the three interferometer mirrors
transmitted about $1$~$\%$ because their high-reflectivity coatings
were not optimized for the angles of incidence applied.
 
\begin{figure}[h]
  \includegraphics[height=8.6cm,angle=-90]{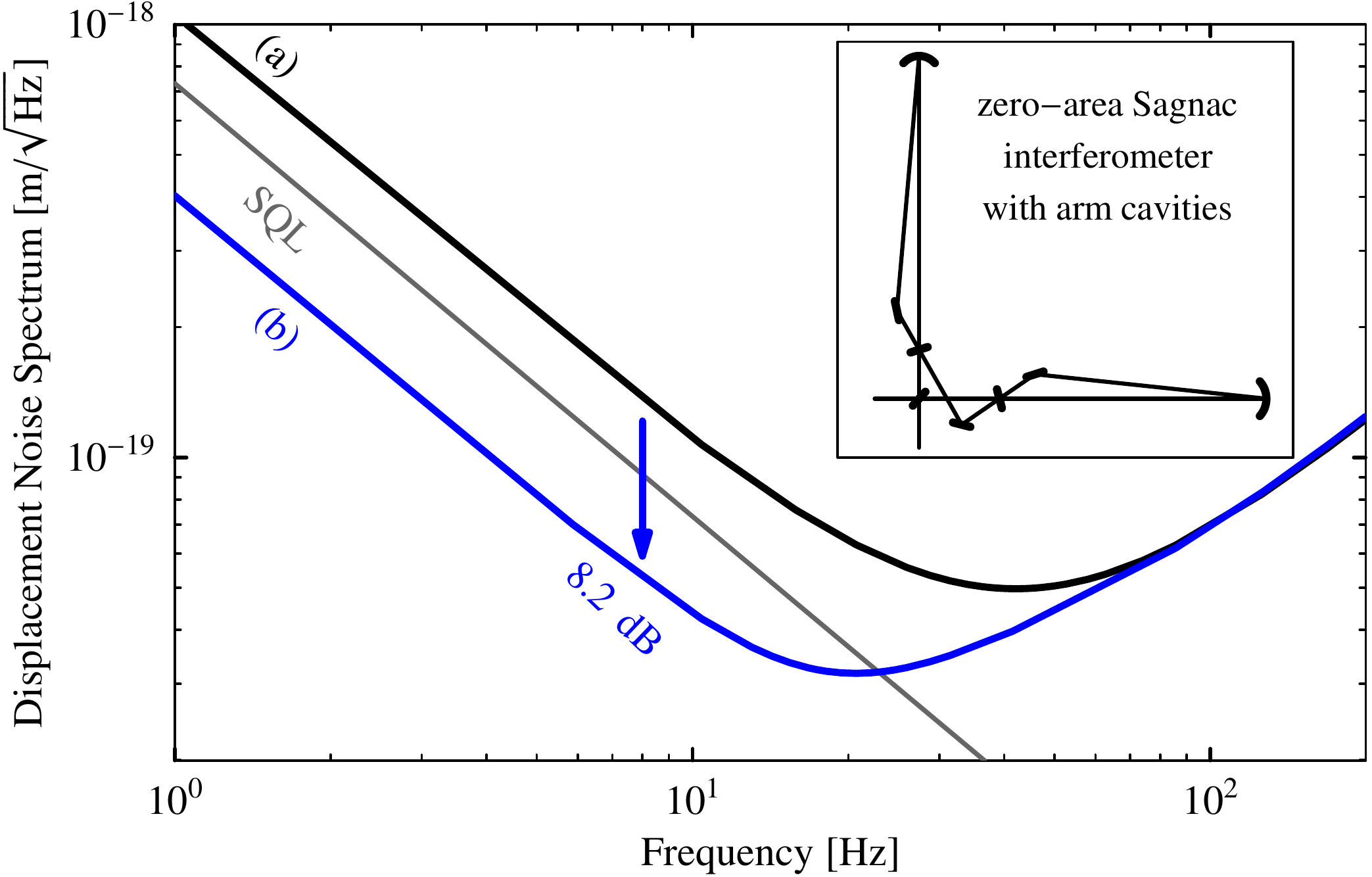}
  \caption{Total quantum noise for a displacement measurement of a
  zero-area Sagnac interferometer with (lower trace) and without (upper trace)
  squeezed light injection. Our model assumes optical losses to the squeezed light of $9$~$\%$ in total, keeping $8.2$~dB of detected squeezing similar to our experiment. The model is further based upon 40\,kg test mass mirrors, and arm-cavities of 10\,km length, 
 80\,Hz linewidth and 10\,kW intra-cavity power. Dividing the linear noise spectral density
  on the Y-axis by 10\,km yields the noise spectral density for a gravitational wave strain
  measurement.}
  \label{fig:ET}
\end{figure}

Figure~\ref{fig:ET} presents the calculated quantum noise spectral densities of a $10$~km Sagnac gravitational-wave detector. Traces (a) and (b) represent the quantum noise without and with squeezed light enhancement, respectively. 
The topology of our simulated Sagnac GW detector is almost
identical to the one in our experiment and is shown in the inset of Fig.~\ref{fig:ET}. The only
difference is that our simulation used two $10$~km long ring resonators. These arm resonators increase the light's storage time and are oriented perpendicular to each other, and thereby optimize the Sagnac's sensitivity to the frequency band between $1$~Hz and about $40$~Hz which is of high astro-physical interest \cite{PYu06}. The squeezing effect at these frequencies is in complete analogy to the one in our experiment. Squeezing down to 1\,Hz was demonstrated in \cite{Vahlbruch07}. We note that in Fig.~\ref{fig:ET} the noise spectral density does increase towards lower frequencies not because of remaining back-action noise but due to the decreasing signal transfer function of the Sagnac interferometer at lower frequencies~\cite{Sun1996}. %This effect makes the linear noise spectral density increasing proportional to the standard quantum limit. 
The topology plotted in the inset slightly differs from the one previously proposed in Ref.~\cite{Chen03} in order to keep the effective area of the Sagnac perfectly zero thereby avoiding noise couplings from the Sagnac effect \cite{Freise09}. 
Note that our topology is still rather simple and does not include a signal-recycling resonator~\cite{Mee1988}. Also a power-recycling resonator \cite{DHKHFMW83pr} is not required if high input powers are readily available from an ultra-stable laser source \cite{Frede05}. But most importantly, our squeezed light enhanced zero-area Sagnac interferometer does not require filter cavities in order to gain a broadband sensitivity improvement. %We come back to this point in the second next paragraph.

The quantum noise spectra in Fig.~\ref{fig:ET} (a) and (b) are calculated for $40$~kg test mass mirrors and a laser power of $90$~W at the central beam splitter. The arm resonators have a linewidth of 80\,Hz and store a circulating power of $10$~kW. For the squeezed light generation, injection, and detection (trace b) we assume optical losses similar to those in our experiment, i.e.\,$3$~$\%$ loss inside the crystal of the squeezed light laser (an escape efficiency of $97$~$\%$), $1$~$\%$ loss due to propagation and non-perfect mode-matching, $2$~$\%$ loss per passage through the Faraday rotator, and $1$~$\%$ photo diode inefficiency. We further assume that the loss inside the interferometer is not significant which should be achievable by optimizing the dielectric coatings of the interferometer mirrors. The total admixture of the vacuum state to the initially pure squeezed state therefore is $9$~$\%$, and the total optical loss to the signal is $3$~$\%$. In order to achieve the detected squeezing factor of $8.2$~dB we use an initially pure squeezed state of $12.4$~dB which is actually lower than in our experiment. Consequently, the antisqueezing inside the interferometer is also smaller, just $12.1$~dB. For the interferometer readout we used a balanced homodyne detector as in our experiment. The phase angle between interferometer signal field and local oscillator was set to $13.7^\circ$. For this detection angle a perfect cancellation of back-action noise inside the arm cavity linewidth is achieved. At higher frequencies back-action is over-compensated. However, our interferometer parameters provide an unchanged noise floor. 
The SQL in Fig.~\ref{fig:ET} is not beaten without the injection of squeezed light because of the high linewidths of the arm cavities and the relatively low laser power applied. A low laser power is important in order to ease the cryogenic operation of the mirrors. We note that the quantum noise spectrum in trace (b) can be modified on-line, i.e. during the operation of the interferometer. By changing the phase-angle of the balanced homodyne detector the overall quantum enhancement can be even extended to the frequencies above the half-linewidth of the arm cavities, however, in this case the quantum enhancement at low frequencies reduces accordingly.  

%\section{Conclusion}
In conclusion, we have experimentally shown that a nonclassical reduction of quantum measurement noise in a Sagnac interferometer is possible with squeezed light injection and balanced homodyne readout. The balanced homodyne detector was an essential part of the experiment since it allowed the optimization of the detected quadrature angle. 
Our theoretical analysis has shown that squeezed light input and balanced homodyne detection readout is highly compatible with the Sagnac's intrinsic back-action evading property. Both types of quantum enhancement are essential for future gravitational wave detectors, in particular for the detection of signals in the astro-physically interesting band from $~1$~Hz to $40$~Hz covering more than five octaves. We therefore have considered a zero-area Sagnac interferometer with
$10$~km long arm cavities targeting this band. We have found that in this signal band a perfect and
broadband evasion of back-action noise together with a broadband nonclassical shot-noise reduction is possible if a certain balanced homodyne detection angle is applied. No filter cavities are required. The latter are mandatory in a Michelson interferometer, where not only the back-action noise depends on Fourier frequency but the antisqueezing of the injected squeezed light also rapidly increases the back-action noise \cite{KLMTV01}. The factor by which the standard quantum limit is surpassed only depends on the (frequency independent) squeezing factor achieved. Low mass mirrors and low laser powers can be used in order to achieve a quantum noise spectral density of the order $3\times10^{-24}$\,Hz$^{-1/2}$. Both, low masses and low powers, are valuable to reduce thermal noise and enable the cryogenic operation of laser interferometers.  A quantum-enhanced zero-area Sgnac interferometer is
therefore a very suitable candidate for the low-frequency part of a future gravitational wave observatory.

%\section{Acknowledgements}
\begin{acknowledgments}
We thank Matthew Reagor for his help on the experiment. This research was supported by the Centre for Quantum Engineering and Space-Time Research, QUEST. H~M-E acknowledges funding from the European Community's Seventh Framework Programme (FP7/2007-2013) under grant agreement N. 211743 (Einstein Telescope Design Study).
\end{acknowledgments}


\begin{thebibliography}{99}
\bibitem{SHOEMAKER2004} B.\,Abbott {\it et~al.}, Nucl.~Instrum.~Methods Phys.~Res., Sect.~A {\bf 517}, 154 (2004).
\bibitem{ACERNES2002} F.\,Acernes {\it et~al.}, Class.~Quant.~Grav. {\bf 19}, 1421 (2002).
\bibitem{WIL2002} B.\,Willke {\it et~al.}, Class.~Quantum~Grav. {\bf 19}, 1377 (2002).
\bibitem{ANDO2001} M.\,Ando {\it et~al.} (TAMA~Collaboration), Phys.~Rev.~Lett. {\bf 86}, 3950 (2001).
\bibitem{ET2009} M. Punturo \textit{et al.}, Class. Quantum Grav. \textbf{27} 084007, (2010); \texttt{http://www.et-gw.eu}.
\bibitem{Caves81} C.\,M.\,Caves, Phys. Rev. D \textbf{23}, 1693 (1981).
\bibitem{McKenzie2002} K.\,McKenzie, D.\,A.\,Shaddock, D.\,E.\,McClelland, Phys. Rev. Lett. \textbf{88}, 231102 (2002).
\bibitem{Vahlbruch2005} H.\,Vahlbruch, S.\,Chelkowski, B.\,Hage, A.\,Franzen, K.\,Danzmann, R.\,Schnabel, Phys. Rev. Lett. \textbf{95}, 211102 (2005).
\bibitem{Goda2008}K.\,Goda \textit{et al.}, %, O.\,Miyakawa, E.\,E.\,Mikhailov, S.\,Saraf, R.\,Adhikari, K.\,McKenzie, R.\,Ward, S.\,Vass, A.\,J.\,Weinstein, N.\,Mavalvala, 
Nat. Phys. \textbf{4}, 472 (2008).
\bibitem{KLMTV01} H.\,J.\,Kimble, %Y.\,Levin, A.\,B.\,Matsko, K.\,S.\,Thorne, S.\,P.\,Vyatchanin, 
\textit{et al.}, Phys. Rev. D {\bf 65} 022002 (2001).
\bibitem{Sagnac1913} G.\,Sagnac, C.R. Hebd. Seances Acad. Sci. \textbf{157}, 708 (1913).
\bibitem{Sun1996} K.-X.\, Sun, M.\,M.\,Fejer, E.\,Gustafson, R.\,L.\,Byer, Phys. Rev. Lett. \textbf{76}, 3053 (1996).
\bibitem{Shaddock1998} D.\,A.\,Shaddock, M.\,B.\,Gray, D.\,E.\,McClelland, Appl. Opt. \textbf{37}, 7995 (1998).
\bibitem{Chen03} Y.\,Chen, Phys. Rev. D. \textbf{67}, 122004 (2003).
\bibitem{Vahlbruch08} H.\,Vahlbruch \textit{et al.}%, M.\,Mehmet, S.\,Chelkowski, B.\,Hage, A.\,Franzen, N.\,Lastzka, S.\,Go{\ss}ler, K.\,Danzmann, R.\,Schnabel
, Phys. Rev. Lett. \textbf{100}, 033602 (2008).
\bibitem{Moritz11dB} M. Mehmet, H. Vahlbruch, N. Lastzka, K. Danzmann, R. Schnabel, Phys. Rev. A \textbf{81}, 013814 (2010).
\bibitem{Takeno2007} Y.\,Takeno, M.\, Yukawa, H.\,Yonezawa, A.\,Furusawa, Opt. Exp. \textbf{15}, 4321 (2007).
\bibitem{PYu06} K. A. Postnov and L. R. Yungelson, Living Rev. Relativity \textbf{9}, 6, (2006).
\bibitem{Vahlbruch07} H.\,Vahlbruch, S.\,Chelkowski, K.\,Danzmann, R.\,Schnabel,  New Journal of Physics 9, \textbf{371}, (2007).
\bibitem{Freise09} A. Freise %, S. Chelkowski, S. Hild, W. Del Pozzo, A. Perreca, and A. Vecchio, 
\textit{et al.}, Class. Quantum Grav. \textbf{26}, 085012 (2009).
\bibitem{Mee1988} B.\,J.\,Meers, Phys.~Rev.~D {\bf 38}, 2317 (1988).
\bibitem{DHKHFMW83pr} R.\,W.\,P.\,Drever {\it et al.}, in  {\it Quantum Optics, Experimental Gravitation, and Measurement Theory}, edited by P.\,Meystre and M.\,O.\,Scully (Plenum, New York, 1983), pp.\,503--514.
\bibitem{Frede05} M. Frede \textit{et al.}%, R. Wilhelm, D. Kracht, and C. Fallnich
, Opt. Express \textbf{13}, 7516 (2005).
%\bibitem{Weiss1986} R.\,Weiss, in NSF Proposal, 1987
%\bibitem{Hild2010} S.\,Hild, S.\,Chelkowski, A.\,Freise, J.\,Franc, N.\,Morgado, R.\,Flaminio, R.\,DeSalvo, Class. Quant. Grav. {\bf 27}, 015003 (2010).

\end{thebibliography}
\end{document}